\documentclass[aps,pra,twocolumn,superscriptaddress,notitlepage,nofootinbib,longbibliography]{revtex4}
\usepackage{bbding}
\usepackage[utf8]{inputenc}
\usepackage{amsthm}
\usepackage{dcolumn}
\usepackage{bm}
\usepackage{subfigure}
\usepackage{epsfig,graphicx,times}
\usepackage{amstext}
\usepackage{amsmath}
\usepackage{amssymb}
\usepackage{graphicx}
\usepackage{latexsym}
\usepackage{bm}
\usepackage{graphicx,epstopdf}
\usepackage[dvipsnames]{xcolor}
\usepackage{epsfig}
\usepackage{graphics}
\usepackage{tabularx}
\usepackage{subfigure}
\usepackage[mathlines]{lineno}% Enable numbering of text and display math
\usepackage{color}
\begin{document}

\title{Enhanced entanglement and controlling quantum steering in a
Laguerre-Gaussian cavity optomechanical system with two rotating mirrors}
\author{Amjad Sohail}
\email{amjadsohail@gcuf.edu.pk}
\affiliation{Department of Physics, Government College University, Allama Iqbal Road,
Faisalabad 38000, Pakistan}
\author{Zaheer Abbas}
\affiliation{Department of Physics, Government College University, Allama Iqbal Road,
Faisalabad 38000, Pakistan}
\author{Rizwan Ahmed}
\affiliation{Physics Division, Pakistan Institute of Nuclear Science and Technology
(PINSTECH), P. O. Nilore, Islamabad 45650, Pakistan}
\author{Aamir Shahzad}
\affiliation{Department of Physics, Government College University, Allama Iqbal Road,
Faisalabad 38000, Pakistan}
\author{Naeem Akhtar}
\affiliation{Department of Physics, Zhejiang Normal University, Jinhua 321004, China}
\author{Jia-Xing Peng}
\affiliation{State Key Laboratory of Precision Spectroscopy, Quantum Institute for Light
and Atoms, Department of Physics, East China Normal University, Shanghai
200062, China}

\begin{abstract}
{Gaussian quantum steering is a type of quantum correlation in which two
entangled states exhibit asymmetry. We present an efficient theoretical
scheme for controlling quantum steering and enhancing entanglement in a
Laguerre-Gaussian (LG) rotating cavity optomechanical system with an optical
parametric amplifier (OPA) driven by coherent light. The numerical
simulation results show that manipulating system parameters such as
parametric gain $\chi$, parametric phase $\theta$, and rotating mirror
frequency, among others, significantly improves mirror-mirror and
mirror-cavity entanglement. In addition to bipartite entanglement, we
achieve mirror-cavity-mirror tripartite entanglement. Another intriguing
discovery is the control of quantum steering, for which we obtained several
results by investigating it for various system parameters. We show that the
steering directivity is primarily determined by the frequency of two
rotating mirrors. Furthermore, for two rotating mirrors, quantum steering is
found to be asymmetric both one-way and two-way. As a result, we can assert
that the current proposal may help in the understanding of non-local
correlations and entanglement verification tasks.}
\end{abstract}

\maketitle

\section{Introduction}

Quantum entanglement is one of the most weirdest phenomenon of quantum
mechanics in which there exists a non-classical (non-local) correlation
between spatially separated states \cite{ent,ent1}. Entanglement has been
shown to have numerous applications in quantum information processing \cite%
{tele,ram}, quantum teleportation \cite{tele,ram}, quantum algorithms \cite%
{algo}, and quantum computing \cite{ent2,comp}. In recent years, many of the
theoretical and experimental studies focused on the generation of
microscopic entanglement in systems such as atoms, ions, and photons \cite%
{micro,micro1,micro2}. However, it remained a pipe dream for researchers to
generate quantum entanglement for macroscopic objects. As a result of hard
efforts, macroscopic entanglement has been experimetally demonstrated for
superconducting qubits \cite{sup}, atomic ensembles \cite{atom} and
entanglement between diamonds \cite{diam}.

In addition to quantum entanglement, Einstein-Podolsky-Rosen (EPR) steering
\cite{epr} is one of an important quantum correlation that sits between
Bell's non-locality and entanglement \cite{bell,bell1,bell2}. As there
exists non-exchangeable asymmetric roles between the two observers, Alice
and Bob. This phenomenon was first introduced by Schr\"{o}dinger to show
that the non-locality in EPR states \cite{sch,sch1}. Most distinctive and
unique feature of quantum steering is its directionality \cite{dir}. It is
interesting to mention here that the steerable states of massive and
macroscopic objects can be utilized to test the foundational issues of
quantum mechanics and the implementation of quantum information processing.
In recent years, many of the studies has focused to achieve these goals in
many optomechanical \cite{stee,stee1,stee2} and in magnomechanical systems
\cite{opent3,opent4,opent5}. It is because of advancement in present day
technology that the most promising feature of quantum steering, one-way
steering, has been experimentally demonstrated \cite{exp,exp1}.

Over the course of about last two decade, cavity optomechanical systems \cite%
{oms,oms2,oms3} have gained importance due to its academic significance and
applications in foundational issues of quantum mechanics \cite{found} and
highly precise measurements \cite{measure}. Optomechanical systems have
recently sparked widespread interest in their discussion of quantumness \cite%
{wig1} via the Wigner function \cite{naeem1,naeem2} and time-frequency
analysis \cite{naeem3}. In addition to above, optomechanical systems have
been experimentally used because of the development of optical micro-cavity
technology. They provide an excellent test bed macroscopic entangled state
\cite{opent,opent1,opent2}, coherent state \cite{SigJS,coh}, squeezed state
\cite{sq,Sin}, ground state cooling \cite{cool} and optomechanically induced
transparency (OMIT) \cite{omit,omit1,omit2,Sn1,Sn2}. In principle, nothing
from the principles of quantum mechanics prohibit macroscopic entanglement.
This has been first investigated by Vitali and collaborators in their
seminal paper \cite{oms}. Therefore, optomechanics is paving a way towards
the macroscopic non-local correlations at macroscopic scales.

Motivated by the progress in optmechanical systems, Bhattacharya and Meystre proposed an optomechanical system comprising a rotating mirror (or rovibrational mirror) which is directly coupled to a Laguerre-Gaussian (LG) cavity mode via exchange of angular momentum \cite{bhatt,bht}. In their scheme, they demonstrate that the rotating mirror can be sufficiently cooled
to 8mK right from the room temperature (300K), due to the incident radiation torque. At such low temperature, one can observe various nonlinear and quantum phenomena. In another interesting scheme, Bhattacharya and collaborators, showed that one can generate entanglement between mechanical rotating mirror and LG cavity mode \cite{bhatt1}. More recently,
researchers found that orbital angular momentum of LG cavity mode can be measured by utilizing OMIT window width \cite{peng,XHYM}. Furthermore, researcher used hybrid rotational system to study OMIT \cite{QingL,peng3,YXW,ZAAK,Rao}, cooling of rotating mirrors \cite{QioP}, and entanglement \cite{peng1,peng2,Singh,FWK}. Most recently, there is an interesting study by Huang \textit{et al}, in which they proposed a LG cavity optomechanical scheme, assisted with an optical parametric amplifier (OPA) whose pump frequency is double the frequency of frequency of anti-Stokes field which is
being generated by the external laser beam interacting with both rotating mirrors, for exploring the entanglement between two rotating mechanical mirrors \cite{huang} and showed that the maximum entanglement is restricted by the gain of the OPA.

The current research focuses on the controllable quantum steering and
enhanced entanglement in a resolved-side-band of two rotating mechanical
mirrors of a single LG optomechanical cavity containing an OPA. We
specifically investigated mirror entanglement as well as quantum steering of
rotating mirrors, and our results show that parametric interactions have\^{A}%
~increased entanglement. Moreover, we also discussed the dependence of
entanglement upon several system parameters such as parametric gain $\chi$,
effective detuning $\Delta$ and parametric phase $\theta$. It is also shown
that there exists tripartitite (mirror-cavity-mirror) entanglement,
quantified by residual cotangle. In addition to entanglement, we also
studied the control of quantum steering $\xi$, which has been found to be
asymmetric both one-way and two-way for rotating mirrors RM$_{1}$ and RM$%
_{2} $, respectively.

The paper is arranged as follows. In Section 2, we present
the theoretical model and the corresponding dynamical quantum langevin
equations. The discussion of methodology for measuring entanglement and
quantum steering is covered in Section 3. We present
numerical results that discuss bipartite/tripartite entanglement and quantum
steering in Section 4 and Section 5,
respectively, then Section 6, shows the experimental
feasibility of the present scheme. Finally, Section 7
presents the conclusions.

%The paper is arranged as follows. In Section \ref{TM}, we present %the theoretical model and the corresponding dynamical quantum langevin equations. The discussion of methodology for measuring entanglement and %quantum steering is covered in Section \ref{ESM}. We present %numerical results that discuss bipartite/tripartite entanglement and quantum %steering in Section \ref{BTE} and Section \ref{CGS}, %respectively, then Section \ref{EI}, shows the experimental %feasibility of the present scheme. Finally, Section \ref{CON} %presents the conclusions.
%Finally, we conclude in \textit{section} in \ref{CON}. %\label{ESM}\label{BTE}\label{CGS}\label{EI}

\section{Theoretical Model}

\label{TM} The system we considered comprises an optical cavity with length $L $ with two rotating mirrors, as shown in \textbf{Figure} 1. The cavity is driven by
the input Gaussian beam which has no topological charges. The Gaussian beam
with frequency $\omega_{L}$ and zero charge enters the cavity through the
left rotating mirror (RM$_{1}$), which is partly transmissive. The reflected
beam gets a charge $-2l$ while the transmitted beam obtain zero charge.
Moreover, the transmitted beam is then incident on the right rotating mirror
(RM$_{2}$) and gets charge $2l$ (zero) on reflection (transmission) from RM$%
_{2}$. Thus a beam with charge $2l$ interacts with the RM$_{1}$ again. The
transmitted beam from RM$_{1}$ has charge $2l$ while the reflected beam
remains with no charge, demonstrating that the net topological charge
increment of light transmission back and forth in the cavity is zero,
ensuring the stability of the system. Moreover, the coupling between the
cavity and left/right rotating mirrors established due to the exchange of
orbital angular momentum. Therefore, we can expect the entanglement between
two rotating mirrors. The exchange of angular momentum between each rotating
mirror and each intracavity photon is $2l\hbar$ and time for a round trip is
$2L/c$, thus the radiation torque acting on both mirror is the change in
angular momentum per the time required for a round trip, that is, $\frac{%
2l\hbar}{(2L/c)}=\frac{cl\hbar}{L}=\hbar g_{\phi}$, where $g_{\phi}=cl/L$
optorotational coupling strength. For simplicity, we assumed the same radius
$R$ and the same mass $m$, and the same intrinsic damping rate $\gamma_{m}$.
In addition, a degenerate OPA is embedded into the cavity. Finally, the beam
is completely reflected from RM$_{2}$. Moreover, we have ignored the loss of
incident light \cite{ASRA}. In a frame rotating at the driving frequency $\omega _{L}$,
the Hamiltonian of the whole system takes the form
%[hbt] The actual placement, [b!] At the bottom, [tbp] At the top.
\begin{figure}[tbp]
\begin{center}
\includegraphics[width=1\columnwidth,height=1.4in]{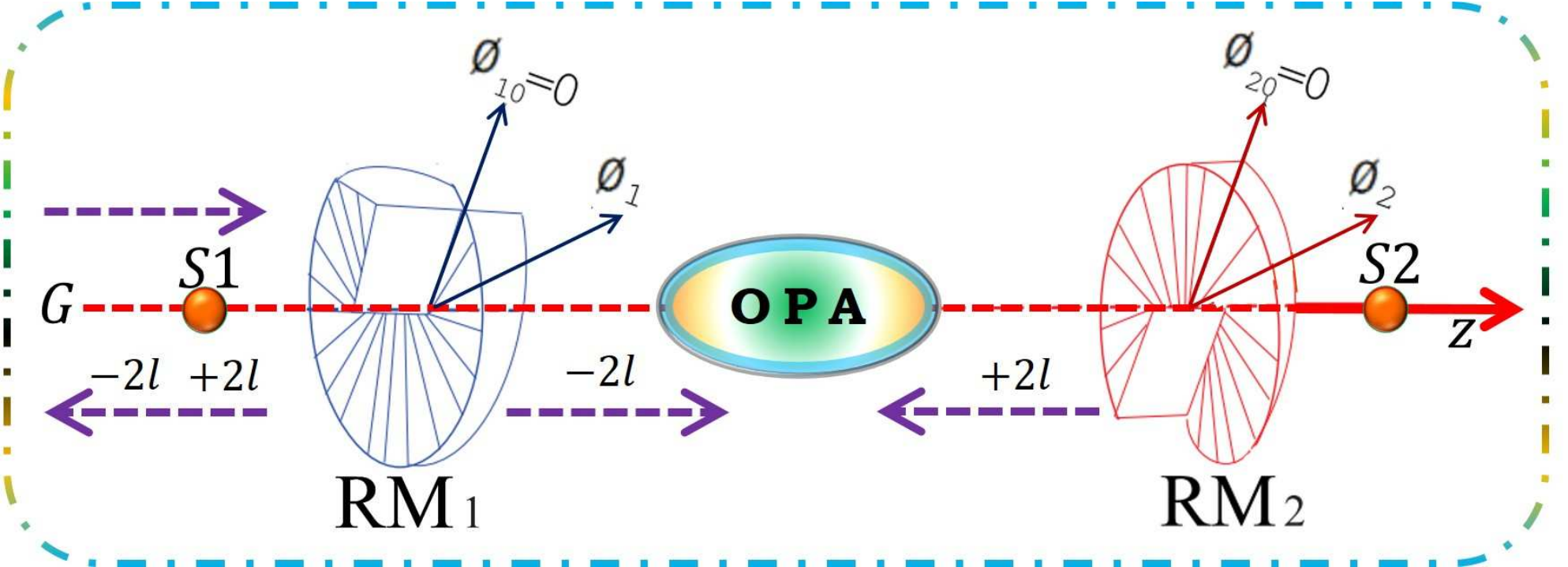}
\end{center}
\caption{Diagrammatic scheme consists of two rotating
mirrors RM$_{1}$ and RM$_{2}$ coupled to the Laguerre-Gaussian cavity mode
via exchange of angular momentum. The two rotating mirrors are mounted on
support S1 and S2, respectively and have the same direction along the cavity
axis $z$. The angular deflections of the two rotating mirror are denoted by
the angle which are taken from their equilibrium position. In addition, a
degenrate OPA is embedded inside the LG-cavity and pumped by a coherent
light.}
\end{figure}
\begin{eqnarray}
H/\hbar &=& \left[\Delta_{0}+(g_{1}\varphi _{1}-g_{2}\varphi_{2})\right]%
a^{\dagger }a+\frac{1}{2}\sum_{j}^{2} \omega_{\varphi _{i}}\left(
L_{z_{i}}^{2}+\varphi _{i}^{2}\right)  \notag \\
&&+i\chi ( a^{\dag ^{2}}e^{i\theta }-a^{2}e^{-i\theta }) +\iota
E\left(a^{\dagger }-a\right),  \label{Eq1}
\end{eqnarray}%
where $\Delta_{0}=\omega _{c}-\omega _{L}$ is the detuning of the cavity
field and $a$\ ($a^{\dagger }$) is the annihilation (creation) operator of
the cavity field, and follow the relation $\left[ a,a^{\dagger }\right] =1$.
While $\phi _{j}$ and $L_{z_{k}}$\ are the angular displacement and angular
momentum of the rotating mirrors, respectively, with $[\phi
_{j},L_{z_{k}}]=i\delta_{jk}(j,k=1,2)$. In above Hamiltonian, the first two
terms are the bare Hamiltonian of the energy for the cavity field and the
two rotating mirrors, the third term describes the optomechanical coupling
between the cavity field and the two rotating mirrors with the coupling
constant given by $g_{i}=(cl/L)\sqrt{\hbar /I\omega _{\phi _{i}}}\left(
i=1,2\right) $ \cite{11}, where $c$ is the velocity of light and $l$ is the
orbital angular momentum quantum number, $L$ is the length of the cavity and
$I=MR^{2}/2$ is the moment of inertia of the two rotating mirrors about the
central axis of the cavity. The next term interprets the interaction of the
2nd-order nonlinear optical crystals, at $2\omega_{L}$, with the L-G cavity
mode with $\chi$ ($\theta$) being the parametric gain (phase) of the OPAs.
The last term of Equation (\ref{Eq1}) represents the interaction of the LG-cavity
mode with the incoming coupling field.

The dynamics of the system can be well described by the time evolution of
the operators and is given by the quantum Langevin equations (QLEs), which
can be written as:
\begin{subequations}
\begin{eqnarray}
\dot{\phi}_{j} &=&\omega _{\phi _{j}}L_{z_{j}},(j=1,2) \label{QLE1}\\
\dot{L}_{z_{j}} &=&-\omega _{\phi _{j}}\phi _{j}+g_{\alpha j}a^{\dagger
}a-\gamma _{m}L_{z_{j}}+\varepsilon ^{in}, \\
\dot{a} &=&-[\kappa +i{\Delta _{\circ }+(g_{1}\phi _{1}-g_{2}\phi _{2})}%
]a+2\chi a^{\dag }e^{i\theta }  \notag \\
&&+E+\sqrt{2\kappa }a^{in},  \label{QLE}
\end{eqnarray}%
where $g_{\alpha {1}}=-g_{1}$, $g_{\alpha {2}}=g_{2}$ and $\Delta _{\circ
}=\omega _{c}-\omega _{L}$. In addition, $\varepsilon ^{in}$ account for the
mechanical noise, and its fluctuation correlations which are relevant to
temperature $T$, can be described by
\end{subequations}
\begin{equation}
\left\langle \delta \varepsilon ^{in}(t)\delta \varepsilon ^{in}(t^{^{\prime
}})\right\rangle =\frac{\gamma _{m_{i}}}{\omega _{\phi _{i}}}%
\int\limits_{-\infty }^{\infty }\frac{d\omega }{2\pi }e^{-\iota \omega
(t-t^{^{\prime }})}\omega \left[ 1+\Theta \right] ,
\end{equation}%
where $\Theta =\coth \left( \frac{\hbar \omega }{2k_{B}T}\right) $, with $%
k_{B}$ being the Boltzmann's constant. Similarly, $a^{in}$ is a noise
operator that of the incident laser beam on the optical cavity and whose
delta-correlated fluctuations are
\begin{equation}
\left\langle a^{in}(t)a^{in,\dagger }(t)\right\rangle =\delta (t-t^{^{\prime
}}),
\end{equation}%
To linearize the set of QLEs, i.e.,  Equation (\ref{QLE1}-\ref{QLE}), we use the ansatz by
writing $\tau =\tau _{0}+\delta \tau $, $(\tau =\phi _{j},L_{z_{j}},a)$,
where $\tau _{s}$ ($\delta \tau $) is the steady-state (quantum fluctuation
) of the operators. Under the long-term limit, the steady-state values of
each operator are:
\begin{subequations}
\begin{eqnarray}
L_{z_{j}} &=&0, \\
\phi _{j0} &=&\frac{g_{\alpha {j}}\left\vert a_{0}\right\vert ^{2}}{\omega
_{\phi _{1}}}, \\
a_{0} &=&\frac{(\kappa -i\Delta +2\chi e^{i\theta })}{(\kappa ^{2}+\Delta
^{2}+4\chi ^{2})}E,
\end{eqnarray}%
where $\Delta =\Delta _{0}+\sum_{j}^{2}g_{j}\phi _{j0}$ represents the
effective cavity detuning while $a_{0}$ denotes the effective field
amplitude of cavity modified by OPA. Since, $|a_{s}|\gg \delta a$, one can
omit the unwanted nonlinear terms, e.g., $\delta a\delta a^{\dagger }$. In
this case, the linearized set of QLEs takes the form: %\begin{eqnarray}
%\delta \dot{\phi}_{1} &=&\omega _{\phi _{1}}\delta L_{z_{1}},  \notag \\
%\delta \dot{L}_{z_{1}} &=&-\omega _{\varphi _{1}}\delta \phi _{1}-\gamma
%_{m}\delta L_{z_{1}}-G_{1}\delta X+\delta \varepsilon ^{in},  \notag \\
%\delta \dot{\phi}_{2} &=&\omega _{\phi _{2}}\delta L_{z_{2}},  \notag \\
%\delta \dot{L}_{z_{2}} &=&-\omega _{\varphi _{2}}\delta \phi _{2}-\gamma_{m}\delta L_{z_{2}}+G_{2}\delta X+\delta \varepsilon ^{in},  \notag \\
%\delta \dot{X} &=&-\kappa \delta X+\Delta \delta Y+\sqrt{2\kappa }X^{in},
%\notag \\
%\delta \dot{Y} &=&-\kappa \delta Y-\Delta \delta X-G_{1}\delta \phi
%_{1}+G_{2}\delta \phi _{2}+\sqrt{2\kappa }Y^{in}, \label{LQE}
%\end{eqnarray}
\end{subequations}
\begin{subequations}
\begin{eqnarray}
\delta \dot{\phi}_{j} &=&\omega _{\phi _{j}}\delta L_{z_{j}}, \label{LQE1}\\
\delta \dot{L}_{z_{j}} &=&-\omega _{\varphi _{j}}\delta \phi _{j}-\gamma
_{m}\delta L_{z_{j}}+G_{j}\delta X+\delta \varepsilon ^{in}, \\
\delta \dot{X} &=&-\kappa \delta X+\Delta \delta Y+\sqrt{2\kappa }X^{in}, \\
\delta \dot{Y} &=&-\kappa \delta Y-\Delta \delta X-G_{1}\delta \phi
_{1}+G_{2}\delta \phi _{2} \notag \\
&&+\sqrt{2\kappa }Y^{in},  \label{LQE}
\end{eqnarray}%
where $G_{j}=\sqrt{2}g_{\alpha j}a_{0}$ is the effective optomechanical
coupling strength. Here, we have introduced the set of quadratures for the
cavity field, that is, $\delta X=\frac{1}{\sqrt{2}}(\delta a+\delta a^{\dag
})$, $\delta Y=\frac{1}{\sqrt{2}i}(\delta a-\delta a^{\dag })$, $\delta
X_{in}=\frac{1}{\sqrt{2}}(\delta a_{in}+\delta a_{in}^{\dag })$ and $\delta
Y_{in}=\frac{1}{\sqrt{2}i}(\delta a_{in}-\delta a_{in}^{\dag })$. Hence, we
rewrite the QLEs (Equation (\ref{LQE1}-\ref{LQE})) in a more compact form, as follow
\end{subequations}
\begin{equation}
\mathcal{\dot{Z}}(t)=\mathcal{A}\mathcal{Z}(t)+n(t),
\end{equation}%
where $\mathcal{Z}^{T}=(\delta \phi _{1},\delta L_{z_{1}},\delta \phi
_{2},\delta L_{z_{2}},\delta X,\delta Y)$ is the fluctuation vector and $%
n^{T}(t)=(0,\varepsilon ^{in},0,\varepsilon ^{in},\sqrt{2\kappa }X^{in},%
\sqrt{2\kappa }Y^{in})$ represents the noise vector. Therefore, the drift
matrix of the system is given by
\begin{equation}
\mathcal{A}=\left(
\begin{array}{cccccc}
0 & \omega _{\phi _{1}} & 0 & 0 & 0 & 0 \\
-\omega _{\phi _{1}} & -\gamma _{m} & 0 & 0 & -G_{1} & 0 \\
0 & 0 & 0 & \omega _{\phi _{2}} & 0 & 0 \\
0 & 0 & -\omega _{\phi _{2}} & -\gamma _{m} & G_{2} & 0 \\
0 & 0 & 0 & 0 & \mu _{+} & \rho _{+} \\
-G_{1} & 0 & G_{2} & 0 & \rho _{-} & \mu _{-}%
\end{array}%
\right) ,
\end{equation}%
where $\mu _{\pm }=-\kappa \pm 2\chi \Re(e^{\theta}) $ and $\rho _{\pm }=\pm
\Delta +\Im(e^{\theta})$. The above drift matrix of the system determines the
stability of the optorotational system, is stable and reaches its
steady-state if none of the eigenvalues of the above drift matrix $\mathcal{A%
}$ has a positive real part \cite{RHC}.
\section{Entanglement and Gaussina Steering measurement}
\label{ESM} \textit{Covariance Matrix.} The foremost task is to find the
stability of any system which, in this case, is obtained by applying the
Routh-Hurwitz criterion which impose certain constraints on the system
parameters. Once the constraints for the current optorotational system are
fulfilled, the steady state of the fluctuations must be a Gaussian state.
Since, $\langle\varepsilon ^{in}\rangle=\langle a^{in}\rangle=0$, and in
this situation, quantum correlations of the system can then be characterized
by the covariance matrix, which has the same order as the drift matrix, with
$V_{jk}=[\mathcal{Z}_{j}(\infty )\mathcal{Z}_{k}(\infty )+\mathcal{Z}%
_{k}(\infty )\mathcal{Z}_{j}(\infty )]/2$, where
\begin{eqnarray}
\mathcal{Z}^{T}(\infty )&=&(\delta \phi _{1}(\infty ),\delta L_{z1}(\infty
),\delta \phi _{2}(\infty ),\delta L_{z2}(\infty ),  \notag \\
&& \delta X(\infty ),\delta Y(\infty ))
\end{eqnarray}
represents the fluctuation matrix. The elements of the covariance matrix
take the form \cite{CWSP}
\begin{equation}
V_{jk}=\sum_{k,l}\int_{0}^{\infty}dt\int_{0}^{\infty}dt^{\prime}%
\Lambda_{i,k}(t)\Lambda_{j,l}(t^{\prime})\Psi_{k,l}(t-t^{\prime}),
\label{VE}
\end{equation}
where $\Lambda=\text{exp}(\mathcal{A}t)$ and
\begin{equation}
\Psi_{k,l}(t-t^{\prime})=\langle
n_{k}(t)n_{l}(t^{\prime})n_{l}(t^{\prime})n_{k}(t)/2
\end{equation}
is noise correlation functions matrix. For large mechanical quality factor,
the quantum Brownian noise can be delta-correlated, that is
\begin{eqnarray}
\Psi_{k,l}(t-t^{\prime})&=&D_{k,l}\delta(t-t^{\prime}),  \notag \\
&=&\text{Diag}[0,\gamma _{m_{1}}(2n_{1}+1),0,\gamma_{m_{1}}(2n_{2}+1),\kappa,\kappa], \notag
\end{eqnarray}
which simplifies Equation (\ref{VE}) as follow
\begin{equation}
V=\int_{0}^{\infty}dt\Lambda(t)D\Lambda^{T}(t).
\end{equation}
The covariance matrix $V$ of system satisfy steady-state Lypunov equation,
\begin{equation}
\mathcal{A}V+\mathcal{A}V^{T}=-D.
\end{equation}
The covariance matrix $V$ which delineate the entanglement configuration of
the current two mirrors optorotational system, its form is
\begin{equation}
V=\left(
\begin{array}{ccc}
M_{1} & N_{12} & N_{1m} \\
N_{12}^{T} & M_{2} & N_{2m} \\
N_{1m}^{T} & N_{2m}^{T} & M_{m}%
\end{array}%
\right) .
\end{equation}
Each entry of above covariance matric $V$ is a $2\times2$ block matrix. The
diagonal entries of the above covariance matrix $V$ describe the local
properties of the participants while the rest of the entries define the the
correlations between different modes. \newline
\textit{Bipartite entanglement.} The logarithmic energy which quantify the
entanglement is defined as
\begin{equation}
E_{N}=\max [0,\ln (2\nu^{-})],  \label{LN}
\end{equation}
where $\nu ^{-}=2^{-\frac{1}{2}}\sqrt{\aleph -\sqrt{\aleph^{2}-4\det
\varsigma}}$ is the smallest symplectic eigenvalue of the partially
transposed covariance matrix (CM), $\varsigma=[\varsigma _{\alpha},\ \varsigma _{\alpha\beta};\varsigma _{\alpha\beta}^{T},\varsigma _{\beta}]$ with each entry of the $\varsigma$ is a $2\times 2$ block matrix and therefore, $\varsigma$ is the $4\times 4$ CM of the bipartite subsystem involving mode $\alpha$ and $\beta$ and
$\aleph =\det \varsigma _{\alpha}+\det \varsigma _{\beta}-2\det \varsigma_{\alpha\beta}$.
%\begin{equation}
%\varsigma =\left(
%\begin{array}{cc}
%\varsigma _{\alpha} & \ \varsigma _{\gamma} \\
%\varsigma _{\gamma}^{T} & \varsigma _{\beta}%
%\end{array}%
%\right) ,
%\end{equation}%
%and
%\begin{equation}
%\aleph =\det \varsigma _{\alpha}+\det \varsigma _{\beta}-2\det \varsigma
%_{\gamma}.
%\end{equation}%
%In addition, each entry of the matrix $\varsigma$ is a $2\times 2$ block
%matrx.

\textit{Tripartite entanglement.} We opt the minimum residual cotangle \cite%
{Adesso,Adesso2,Coffman} for the basic criteria to realize tripartite
entanglement, is given by
\begin{equation}
\mathcal{R}_{\tau }^{min}\equiv \min [\mathcal{R}_{\tau }^{m_{1}|m_{2}a},%
\mathcal{R}_{\tau }^{m_{2}|m_{1}a},\mathcal{R}_{\tau }^{a|m_{1}m_{2}}],
\label{RS}
\end{equation}%
where $\mathcal{R}_{\tau }^{f|jk}\equiv C_{f|jk}-C_{f|j}-C_{f|k}\geq 0$ ($%
f,j,k=m_{1},m_{2},a$) is the monogamy of quantum entanglement and this
condition assures the invariance of tripartite entanglement under possible
permutations of all modes. Hence, $\mathcal{R}_{\tau }^{min}$ and is
describes the genuine tripartite entanglement of any three-mode Gaussian
state. In Equation (\ref{RS}), $C_{f|j}$ is the contangle of subsystems of $f$
and $j$ ( $j$ incorporate one or two modes), and is equal to the \textit{%
squared} logarithmic negativity. To execute \textit{one-mode-vs-two-modes}
logarithmic negativity $E_{f|jk}$, one must reform $\nu^{-}$ (see Equation (\ref%
{LN})) as
\begin{equation}
\nu^{-}=\min eig\ |\bigoplus_{j=1}^{3}(-\sigma _{y})\widetilde{V_{4}}|,
\end{equation}
where $\widetilde{V_{4}}=\mathcal{P}_{1|2}\mathcal{V}_{4}\mathcal{P}_{1|2}$
with $\widetilde{\mathcal{Q}}=\mathcal{G}_{i|jk}\mathcal{Q}\mathcal{G}_{i|jk}$.
Furthermore Furthermore, the partial transposition matrices are given by
\begin{eqnarray*}
\mathcal{G}_{1|23} &=& \text{diag}(1,-1,1,1,1,1), \\
\mathcal{G}_{2|13} &=& \text{diag}(1,1,1,-1,1,1), \\
\mathcal{G}_{3|12} &=& \text{diag}(1,1,1,1,1,-1).
\end{eqnarray*}
%, $\mathcal{G}_{1|23}=$ diag$(1,-1,1,1,1,1)$, $\mathcal{G}%
%_{2|13}= $ diag$(1,1,1,-,1,1)$ and $\mathcal{G}_{3|12}=$ diag$(1,1,1,1,1,-1)$
%are partial transposition matrices.
%In Fig. 6, we show tripartite entanglement among three participants, i.e., magnon-phonon-photon, as shown by the nonzero minimum residual contangle $\mathcal{R}^{min}_{\tau}$
%for different values of $\Delta_{m}$. It has been shown that the maximum tripartite entanglement can be achieved when $\Delta_{m}=0.8\omega_{b}$. This is because the bipartite entanglement $E_{N}^{bm}$ which is much stronger than other biprtitions, is maximum around $\Delta_{m}=0.8\omega_{b}$ as shown in Fig. 2(c). Furthermore, one can notice that by increasing $\Delta_{m}$, there is slight shift of the optimal tripartite entanglement from $\Delta_{c}=-0.6\omega_{b}$ to $\Delta_{c}=-1.2\omega_{b}$. In this way, we can manipulate the system's parameters to obtain optimal entanglement.
\begin{table}%[b!]
\centering
\begin{tabular*}{0.48\textwidth}{@{\extracolsep{\fill}}||l c c|| }
\hline
Parameters & Symbol & Value
\tabularnewline
\hline
Length of the cavity   & $L$ & $1$mm
\tabularnewline

Mass of rotating mirrors  & $m$ & $35$ng
\tabularnewline

Radius of rotating mirrors  & $R$ & $10\mu$m
\tabularnewline

Left mirror angular frequency  & $\omega_{\phi_{1}}/2\pi$ & $ 10^{7}$Hz
\tabularnewline

Right mirror angular frequency  & $\omega_{\phi_{2}}$ & $ (0.5-1.5)\omega_{\phi_{1}}$
\tabularnewline

Input laser power  & $\wp$ & $50$mW
 \tabularnewline

Quality factor & $Q$ & $2\times 10^{7} $
\tabularnewline

Laser wavelength & $\lambda$ & $810nm$
\tabularnewline

Optical finesse  & $F$ & $5\times 10^{3} $
\tabularnewline

Orbital angular momentum  & $l$ & $100\hbar$
\tabularnewline

Temperature  & T & 15 mK
\tabularnewline
\hline
\end{tabular*}
\caption{Experimental values of the  parameters used in the paper \cite{bhatt,bhatt1,peng1}} %\cite{1}.
\label{table:pvalue}
\end{table}
\begin{figure}[b!]
\begin{center}
\includegraphics[width=1\columnwidth,height=2in]{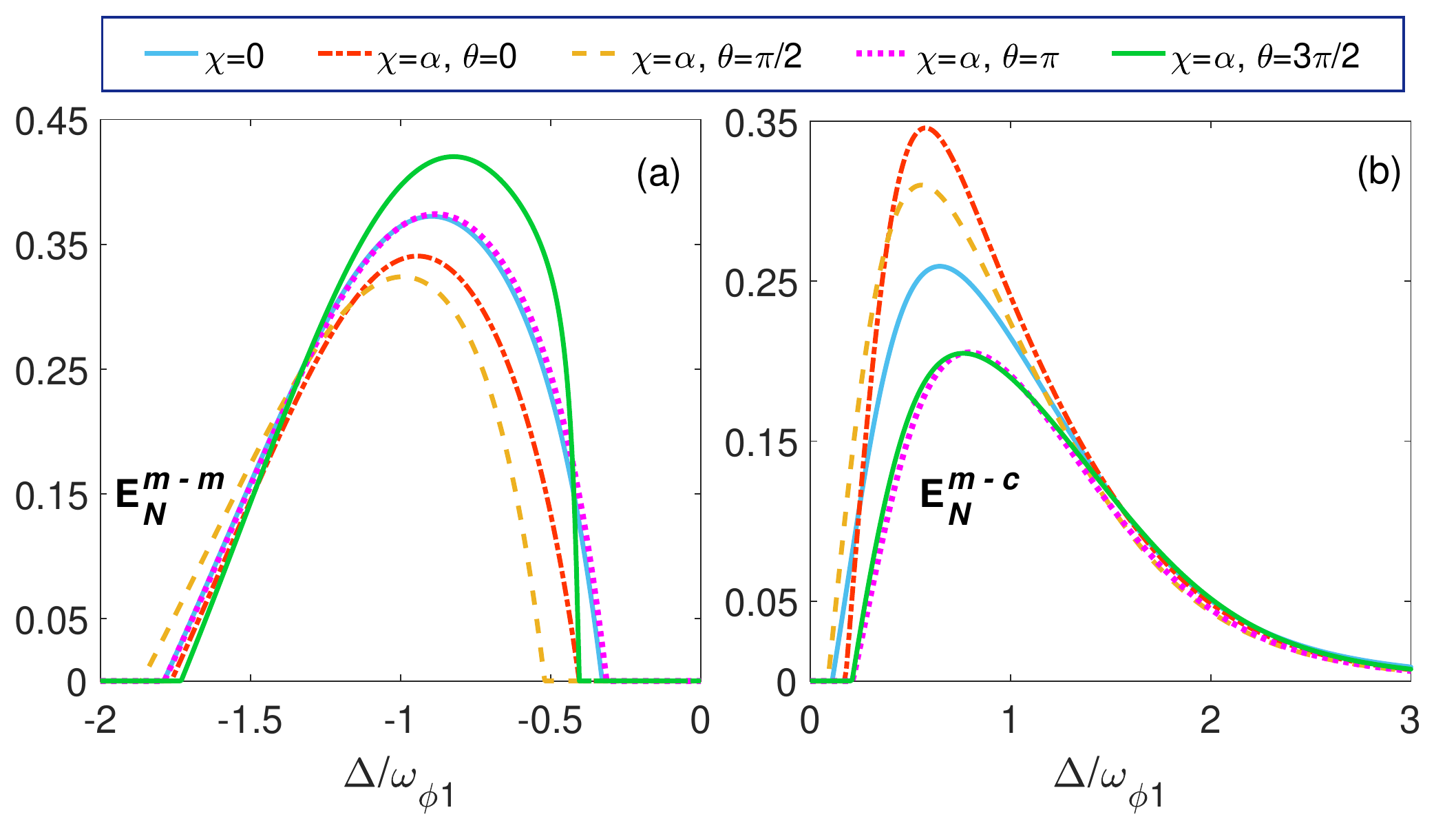}
\end{center}
\caption{(a) The mirror-mirror entanglement $E_{N}^{m-m}$ and
(b) the mirror-cavity entanglement $E_{N}^{m-c}$ plotted as a function of
the normalized detuning $\Delta/\protect\omega_{\protect\phi1}$.}
\end{figure}
\textit{Gaussian steering.} Gaussian steering has asymmetric characteristics
in nature and is much different from the entanglement. For the two
interacting mode Gaussian state, the recommended measurements of the quantum
steerability between any two interacting parties in different directions can
be calculated as
\begin{subequations}
\begin{eqnarray}
\zeta _{\alpha |\beta } &=&\max \{0,\mathcal{S}(2\mathcal{V}_{m_{l}})-%
\mathcal{S}(2\mathcal{V}_{in})\},  \label{A} \\
\zeta _{\beta |\alpha } &=&\max \{0,\mathcal{S}(2\mathcal{V}_{m_{2}})-%
\mathcal{S}(2\mathcal{V}_{in})\},  \label{B}
\end{eqnarray}%
where $\mathcal{S}(\nu )$ is the R\'{e}nyi-2 entropy and is $\mathcal{S}(\nu
)=\frac{1}{2}\ln \det (\nu )$, and
\end{subequations}
\begin{equation}
\mathcal{V}_{in}=\left[
\begin{array}{cc}
\mathcal{V}_{m_{1}} & \mathcal{V}_{m_{1}m_{2}} \\
\mathcal{V}_{m_{1}m_{2}}^{T} & \mathcal{V}_{m_{2}}%
\end{array}%
\right] ,  \label{Co}
\end{equation}%
The diagonal entries of $\mathcal{V}_{m_{1}}$ and $\mathcal{V}_{m_{2}}$
defines the reduced states of modes $\mathcal{V}_{m_{1}}$ and $\mathcal{V}%
_{m_{2}}$, respectively. We quantify the quantum steering by $\zeta _{\alpha
|\beta }$ which translates the mode $\alpha $ steers mode $\beta $ and
similarly $\zeta _{\beta |\alpha }$ defines the swaped direction. It is an
established fact that the entanglement is symmetric property. However, the
EPR steering is intrinsically different from entanglement, i.e., asymmetric
property which means that a quantum state may be steerable from Bob to Alice
but not vice versa. Therefore, there exist mainly three different
possibilities for quantum steering. (i) $\zeta _{\alpha |\beta }=\zeta
_{\beta |\alpha }=0$ as no-way (NW) steering, (ii) $\zeta _{\alpha |\beta }>0
$ and $\zeta _{\beta |\alpha }=0$ or $\zeta _{\alpha |\beta }=0$ $\zeta
_{\beta |\alpha }>0$ as one-way (OW) steering and $\zeta _{\alpha |\beta }>0$
$\zeta _{\beta |\alpha }>0$ as two-way (TW) steering. Asymmetric
steerability between the two modes of the Gaussian states can be checked by
introducing the steering asymmetry, defined as
\begin{equation}
\zeta _{M}=|\zeta _{\alpha |\beta }-\zeta _{\beta |\alpha }|. \label{SAS}
\end{equation}%
It is vital to mentioned here that we obtain one-way or two way steering as
long as steering asymmetry remains positive, i.e., $\zeta _{M}>0$ and no-way
if $\zeta _{M}=0$.
\begin{figure}[tbp]
\begin{center}
\includegraphics[width=0.9\columnwidth,height=2.4in]{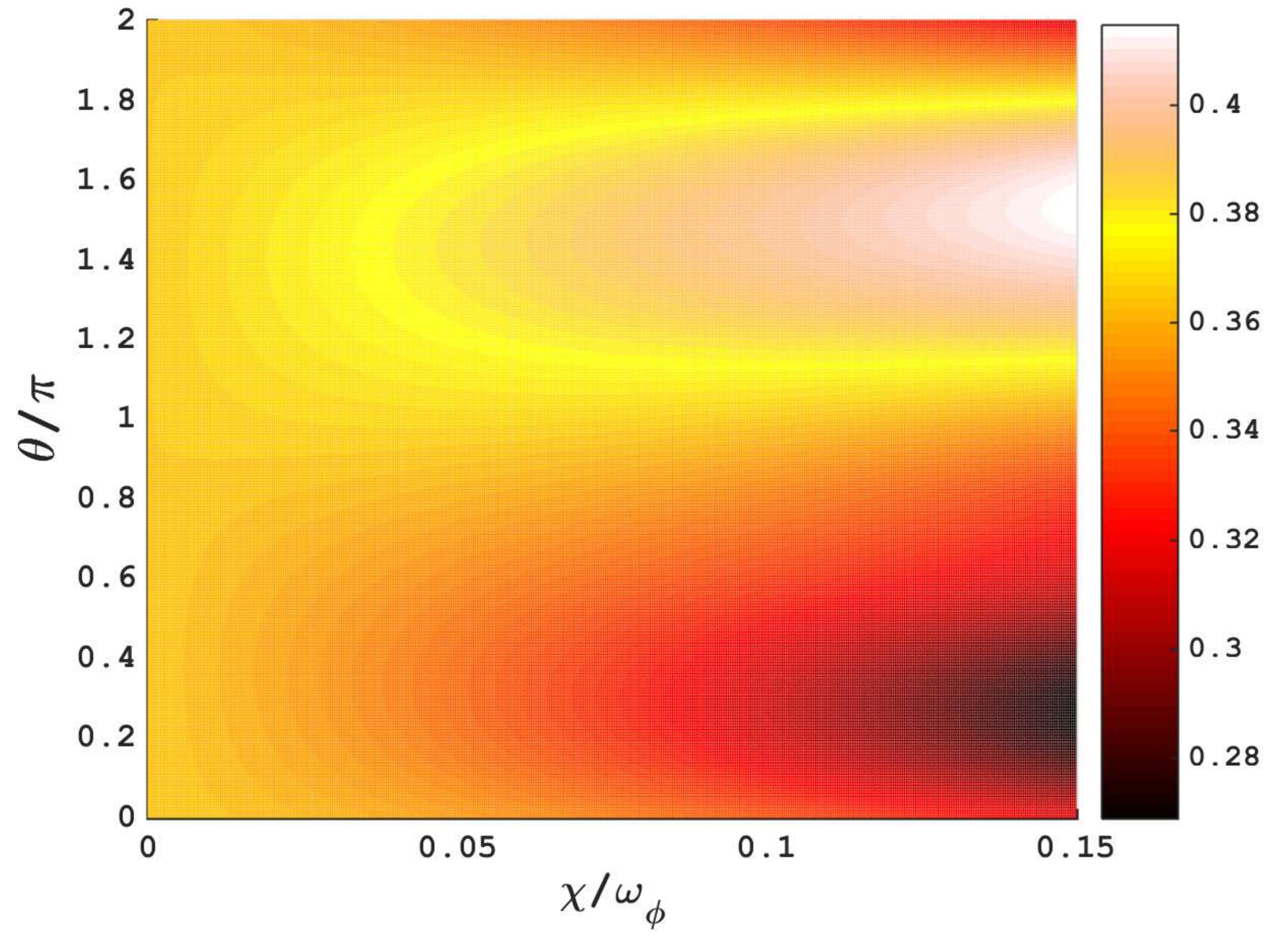}
\end{center}
\caption{(Color Online) Density plot of (a) the mirror-mirror entanglement $%
E_{N}^{m-m}$ versus $\protect\chi $ and $\protect\phi $.}
\end{figure}
\begin{figure}[b]
%[tbp]
\par
\begin{center}
\includegraphics[width=1\columnwidth,height=2.6in]{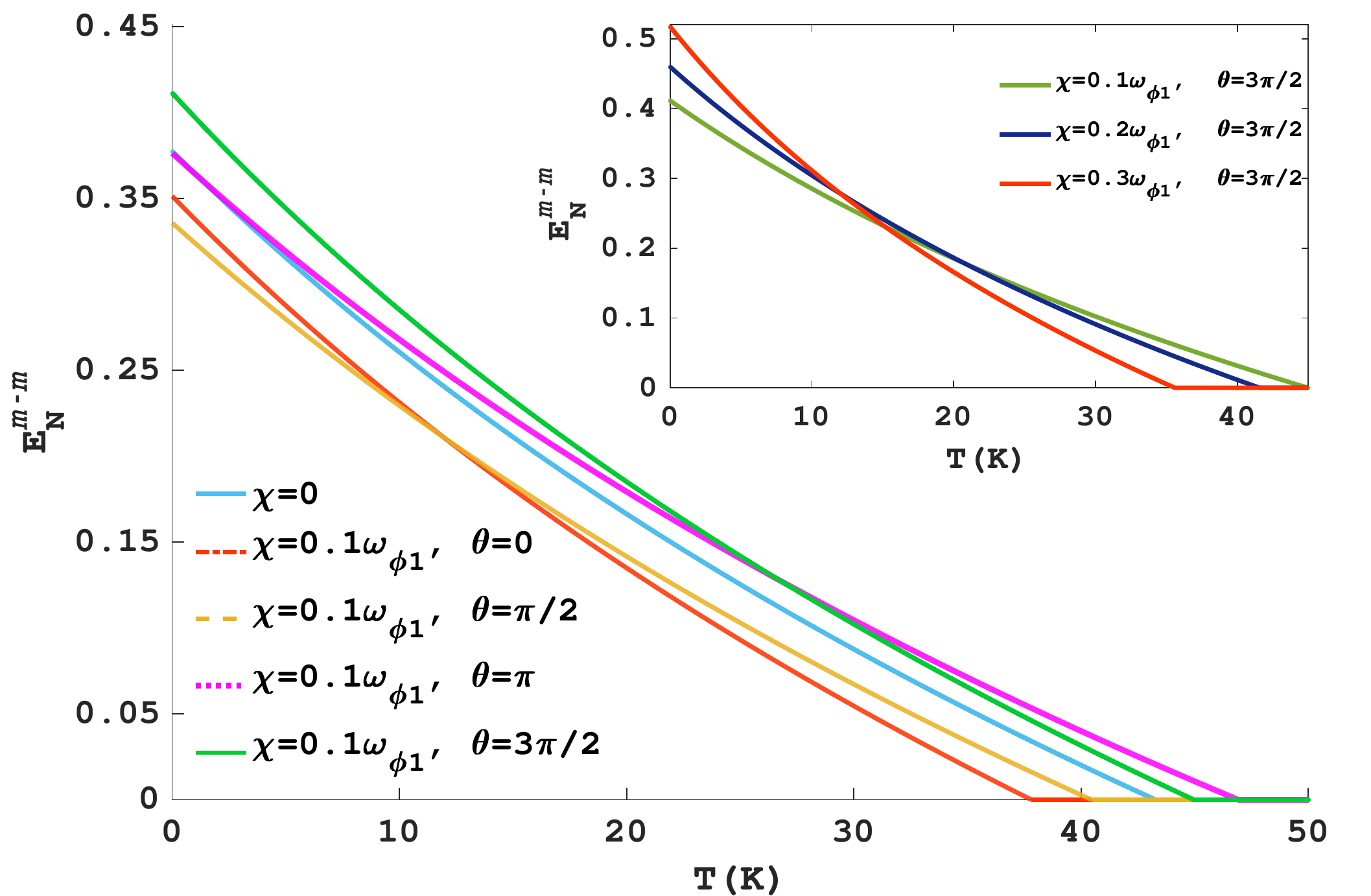}
\end{center}
\par
% $\Delta/\omega_{\protect\phi_{1}}=-1$
\caption{The logarithmic negativity $E_{N}^{m-m}$ as a
function of temperature in the absence and presence of the OPA with optimum
detuning.}
\end{figure}
\begin{figure}[b]
\begin{center}
\includegraphics[width=1\columnwidth,height=2.6in]{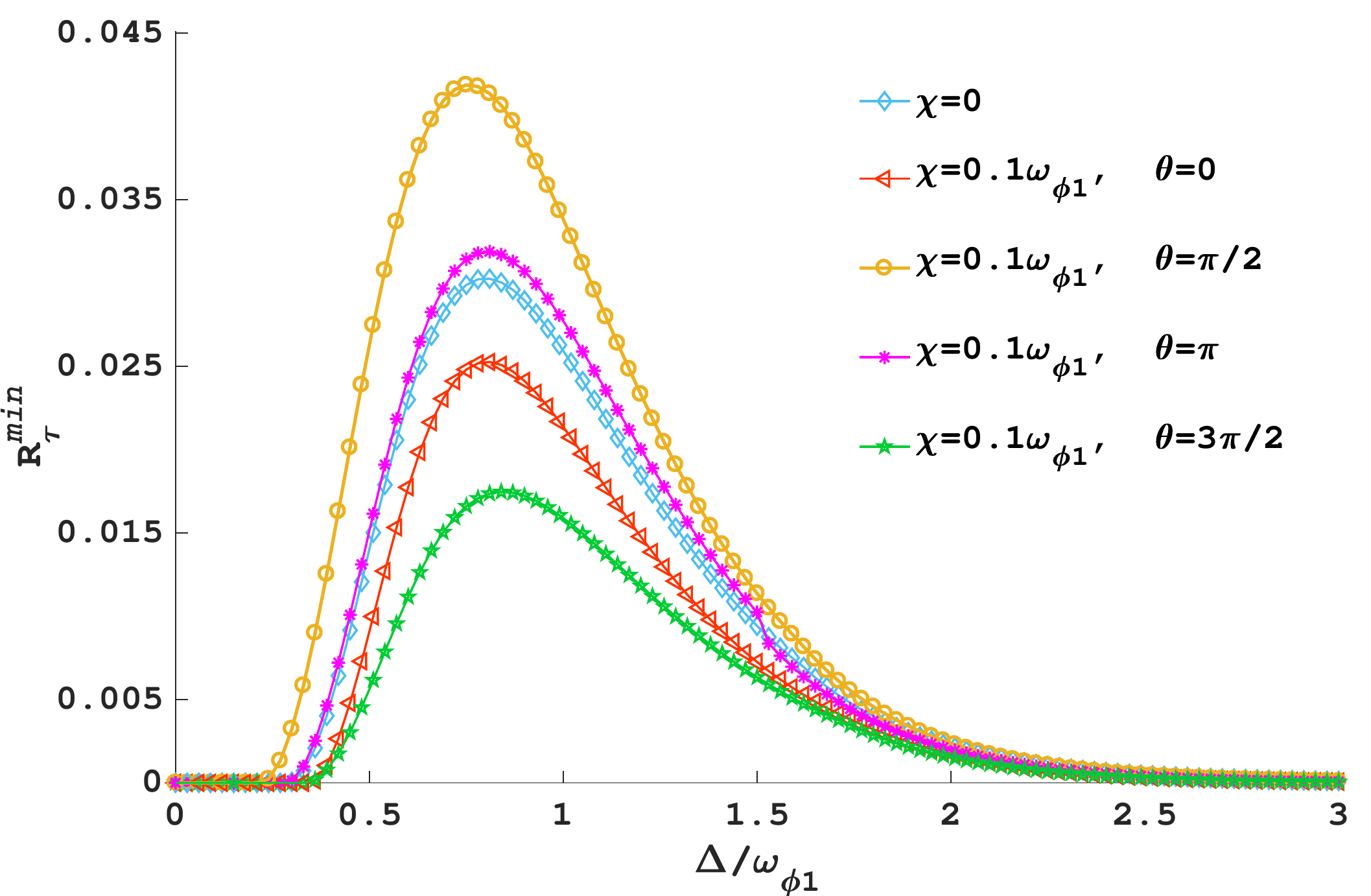}
\end{center}
\caption{Genuine tripartite entanglement
(mirror-cavity-mirror) versus the normalized detuning. The parameters are
same as given in Table.}
\end{figure}

\section{Bipartite and Tripartite Entanglement}

\label{BTE} In the following discussion, we demonstrate the results
regarding the generation of bipartite (and tripartite) entanglement and
quantum steering for the current system. Throughout our numerics, we have
assumed experimentally realizable parameters, similar to those in ref \cite%
{bhatt,bhatt1,peng1}, listed in Table 1. In this paper, we consider two
bi-partitions, namely $E_{N}^{m-m}$ and $E_{N}^{m-c}$ to represent the
degree of entanglement between two phonon modes and between L--G cavity mode
and phonon mode, respectively. The utmost task of studying the entanglement
in such a double rotational cavity optomechanical system is to find optimal
value of the cavity detuning $\Delta$. In \textbf{Figure} 2 (a-b), the logarithmic
negativity is plotted as a function of the detuning in the absence and
presence of parametric gain (with different parametric phases). To ensure
the stability of the system, we choose the parameter gain $%
\chi=0.1\omega_{\phi_{1}}$. We find that the steady state entanglement
between two mirrors $E_{n}^{m-m}$ (between the mirror and the cavity $%
E_{n}^{m-c}$) is obtained when $\Delta<0$ ($\Delta>0$) in the steady state
which is similar to \cite{bhatt1,peng1}. In addition, the maximal value of $%
E_{n}^{m-m}$ ($E_{n}^{m-c}$) is obtained at about $\Delta=-\omega_{\phi_{1}}$
($\Delta=\omega_{\phi_{1}}$). Furthermore, the enhancement pattern of
entanglement $E_{n}^{m-m}$, by varying the phase of the parametric, behaves
almost in converse manner as compared to $E_{n}^{m-c}$ as shown by \textbf{Figure} 2
(a-b).

Next, we examine the combined effect of parametric gain and parametric phase
(associated with the gain), on the mirror-mirror entanglement $E_{n}^{m-m}$
and mirror-cavity entanglement $E_{n}^{m-c}$ as shown by the \textbf{Figure} 3.
Physically, the gradual increase of parametric gain induces the nonlinearity
in the system and consequently, we observe the enhancement in entanglement.
One can see that bipartite entanglement $E_{n}^{m-m}$ increases as the $%
\theta$ increases for a larger value of parametric gain. We can also see
from the \textbf{Figure} 3 that by ensuring the stability of the current system, the
maximum enhancement of $E_{n}^{m-m}$ appears around $\theta=3\pi/2$ for
varied parametric gain. For specific value of $\chi$, the amount of maximum
entanglement increase, is different. For example, keeping the fixed
parametric gain, i.e., $\chi=0.1\omega_{\phi_{1}}$, the amount of maximum
entanglement $E_{n}^{m-m}$ increased, is about $14\%$.

\begin{figure}[tbp]
\begin{center}
\includegraphics[width=1\columnwidth,height=2.3in]{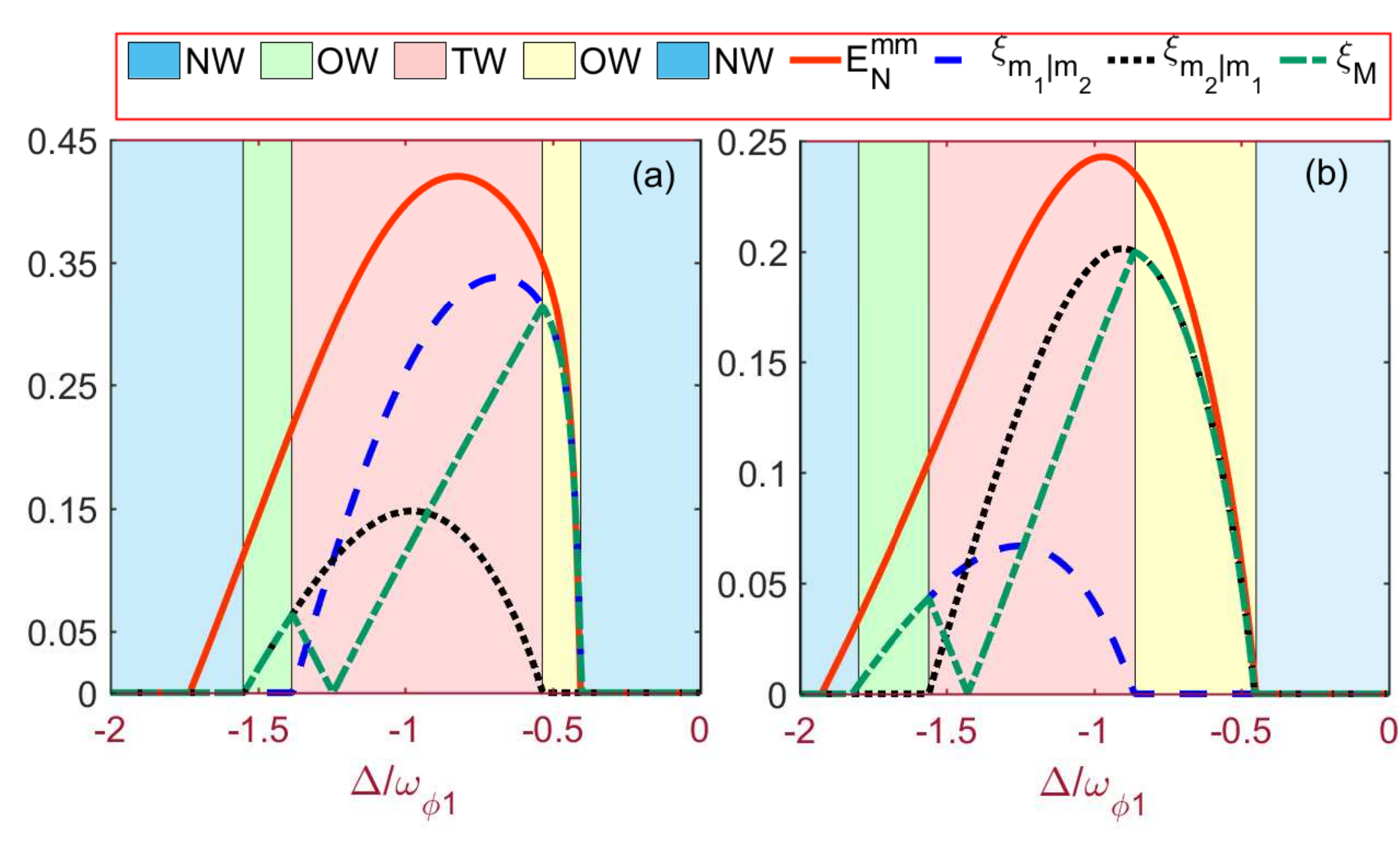}
\end{center}
\caption{Plot of the mirror-mirror entanglement $E_{N}^{m-m}$
(solid orange line), Gaussian steering $\protect\zeta_{m_{1}|m_{2}}$ (dashed
blue line), $\protect\zeta_{m_{2}|m_{1}}$ (dotted black line) and the
steering asymmetry $\protect\zeta_{M}$ (dot-dashed green line) of the two
rotational modes as a function of the normalized detuning $\Delta/\protect%
\omega_{\protect\phi1}$ when (a) $\protect\omega_{2}=0.5\protect\omega_{1}$
and (b) $\protect\omega_{2}=1.5\protect\omega_{1}$. We set $\protect\chi=0$.
The rest of the parameters are same as in Table. 1.}
\end{figure}
The robustness of such a mirror-mirror entanglement in two rotating cavity
optomechanical system with respect to the temperature $T$ as shown in \textbf{Figure}
4. It is clear that for a fixed value of the parametric gain, the amount of
entanglement decreases monotonically as the temperature of the environment
increases. and eventually cease to exist. However, the critical temperature
of mirror-mirror entanglement fluctuate by varying the parametric phase. On
the other hand, the curve shown in the inset of \textbf{Figure} 4 have dissimilar
tendency i.e., by choosing the optimal value of the parametric phase, the
degree of mirror-mirror entanglement $E_{n}^{m-m}$ increases with the rise
of parametric gain. However, the temperature decreases, implying that the
system with larger parametric gain mass possess weaker capability to refrain
from decoherence of the thermal environment.
%the intensity of the mirror-mirror entanglement decreases and finally with the rise of temperature.
%By choosing the optimal value of the cavity detuning which is $\Delta=-1$, if the parametric gain increases, the degree of mirror-mirror entanglement $E_{n}^{m-c}$ increases as shown by the inset in Fig. 4. However, the temperature decreases, which is very beneficial in quantum information processing.
%In Fig. 4, we explore the effect of parametric gain and parametric phase

One of the most vital contribution in the present double rotating system is
to find the possibility of exploring the genuine tripartite entanglement. In
\textbf{Figure} 5, we investigate the genuine tripartite entanglement versus normalized
detuning. It is clear that the genuine tripartite entanglement among the
participants (i.e., phonon-photon-phonon) exists in the current double
rotating system if the cavity mode is in blue sideband. The solid blue curve
shows the genuine tripartite entanglement when $\chi=0$. However, in the
presence of parametric gain $\chi=0.1\omega_{\phi_{1}}$, we plot tripartite
entanglement for different value of phase. It is clear from \textbf{Figure} 5 that
maximum enhancement in tripartite entanglement is obtained for $\theta=\pi/2$%
. It is also noticeable that this enhancement go along with an increase in
domain of genuine tripartite entanglement over a relatively wider range of
effective cavity detuning. Furthermore, the maximum of genuine tripartite
entanglement with parametric phase increases by $40\%$. Hence, the
parametric phase of the OPA plays a vital role in enhancing the genuine
tripartite entanglement as exhibited in \textbf{Figure} 5.

\section{Controlling Gaussian Steering}

\label{CGS}
\begin{figure}[b!]
\begin{center}
\includegraphics[width=1\columnwidth,height=3.5in]{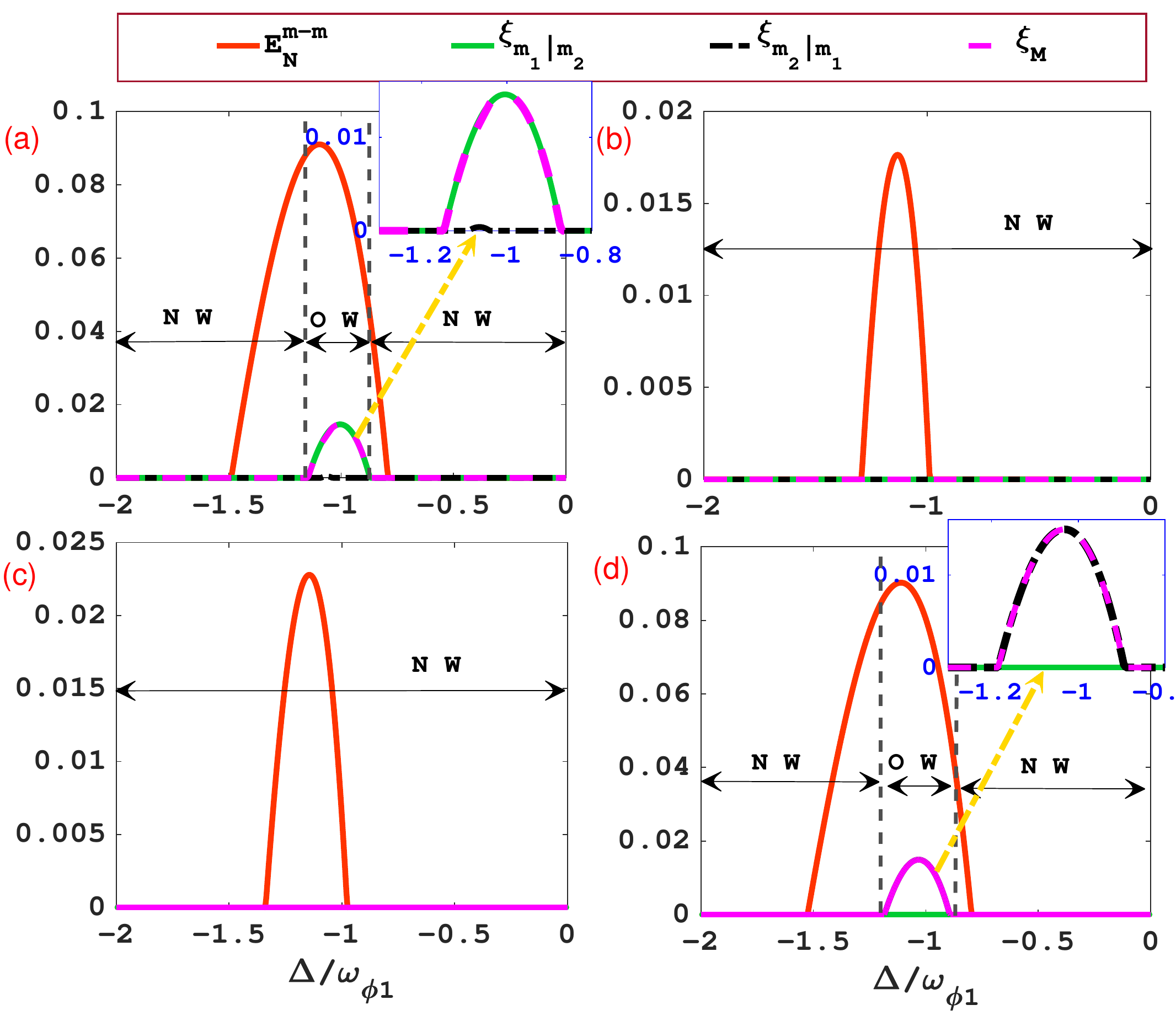}
\end{center}
\caption{The mirror-mirror entanglement (solid orange line), Gaussian steering $\protect\zeta_{m_{1}|m_{2}}$ (solid
green line), $\protect\zeta_{m_{2}|m_{1}}$ (dot-dashed black line) and the
steering asymmetry $\protect\zeta_{M}$ (dashed magenta line) as a function of the normalized detuning $\Delta/\protect%
\omega_{\protect\phi1}$ when (a) $\protect\omega_{\protect\phi_{2}}=0.9%
\protect\omega_{\protect\phi_{1}}$, (b) $\protect\omega_{\protect\phi%
_{2}}=0.95\protect\omega_{\protect\phi_{1}}$, (c) $\protect\omega_{\protect%
\phi_{2}}=1.05\protect\omega_{\protect\phi_{1}}$ and (d) $\protect\omega_{%
\protect\phi_{2}}=1.1\protect\omega_{\protect\phi_{1}}$. The rest of the parameters are same as in Fig. 6 and in Table. 1.}
\end{figure}
\begin{figure*}[tbp]
\begin{center}
\includegraphics[width=2\columnwidth,height=5in]{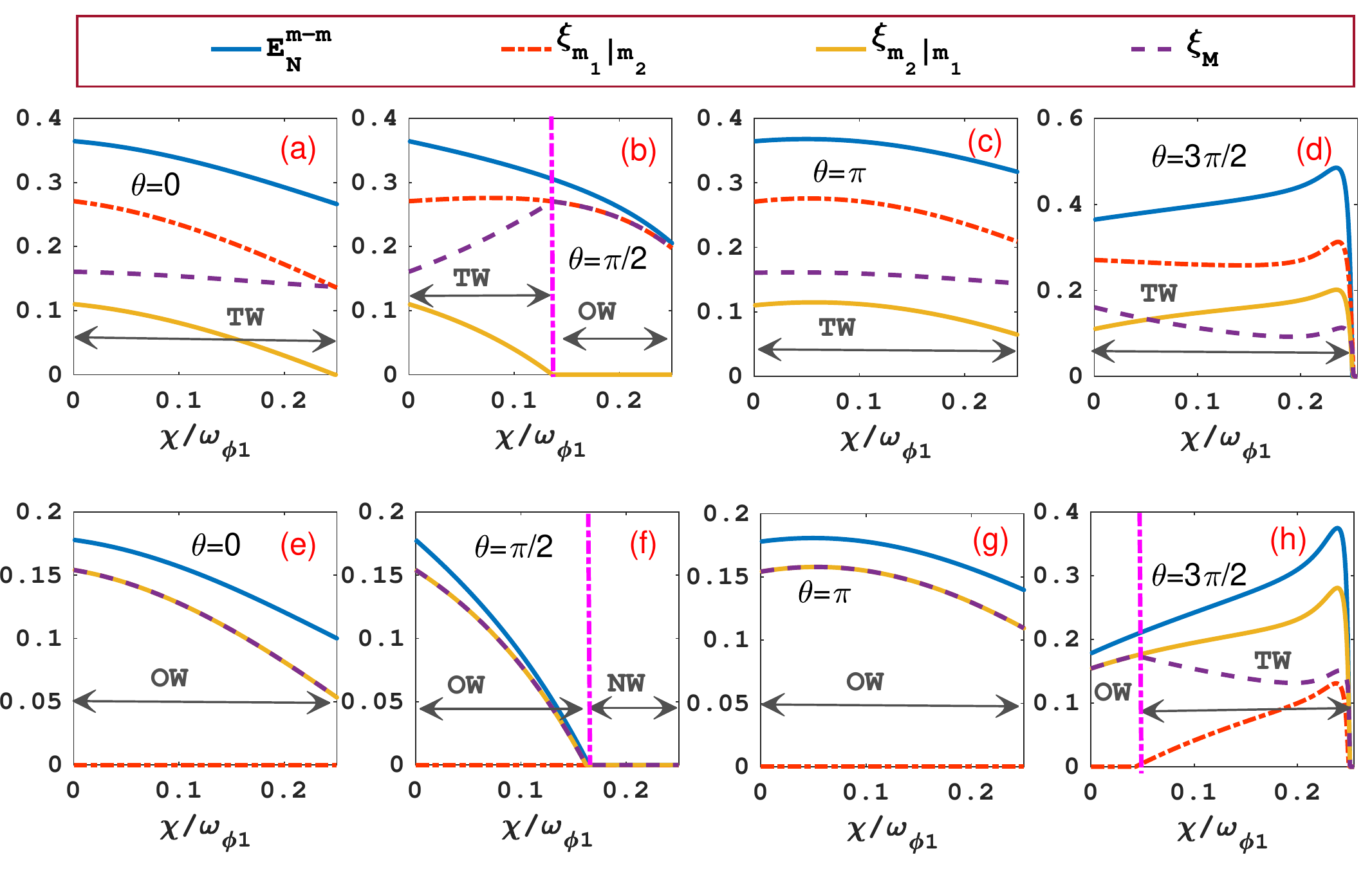}
\end{center}
\caption{Quantum steering as a function of $\protect\chi/%
\protect\omega_{\protect\phi1}$ for different values of $\protect\theta$:
(a)-(e) for $\protect\theta=0$, (b)-(f) is for $\protect\theta=\protect\pi/2$%
, (c)-(g) represents the cases when $\protect\theta=\protect\pi$, and
(d)-(h) with $\protect\theta=3\pi/2$. In addition, we set $%
\protect\omega_{\protect\phi_{2}}=0.5\protect\omega_{\protect\phi_{1}}$ in
(a)-(d) and $\protect\omega_{\protect\phi_{2}}=1.5\protect\omega_{\protect%
\phi_{1}}$ in (e)-(h).We set the optimal value of the detuning.
The rest of the parameters are kept same as given in Table. 1.}
\end{figure*}
\textit{Two-way (TW) control.} In the following, our purpose is to find the
controllable quantum steering in the double rotational cavity optomechanical
system. Since the entanglement and the quantum steering are two different
facets of inseparable quantum correlations, therefore, it is better to study
them simultaneously under the same conditions. For this purpose, we employ
the R\'{e}nyi-2 entropy to quantify the steerability between the two
rotational modes. We first demonstrate this idea in \textbf{Figure} 6, the peak values
of the entanglement and quantum steering of the two mechanical modes. It is
evident from \textbf{Figure} 6. that steerable states are always entangled, however,
the reverse order is not generally true, i.e., entangled states are not
usually steerable. This property reflects the idea that stronger quantum
correlations, between the two rotational modes, are indispensable for
realizing the Gaussian steering than that for the entanglement. In such a
two-rotating cavity optomechanical system, there may exist (one-) two-way
quantum steering between the two rotational modes due to frequency
difference. To demonstrate the Gaussian steering more clearly, we use the
pink area to area to depict the presence of two-way steering, green and
yellow areas for one-way steering, and the blue area to represent the
presence of no-way steering. It is clear from \textbf{Figure} 6 (a-b), that both the
mirror steer each other, however, the steerability mainly depends on the
frequency of each mirror. The mirror which has high frequency steers more as
compared to the mirror which has less frequency as clearly demonstrated by
the pink area in \textbf{Figure} 6 (a-b). The two-way steering implies that both Alice
and Bob can convince each other that the state they shared, is entangled.
Furthermore, \textbf{Figure} 6. reveal situation where $\zeta_{m_{1}|m_{2}}>0$, $%
\zeta_{m_{2}|m_{1}}=0$ and $E_{N}^{m-m}>0$ (or $\zeta_{m_{1}|m_{2}}=0$, $%
\zeta_{m_{2}|m_{1}}>0$ and $E_{N}^{m-m}>0$) witnesses the Gaussian one-way
(OW) steering region as shown by the green (yellow) region. One-way (OW)
steering in the current system translates that the states of the two
rotational modes are steerable in single direction even though they are
entangled. Such a kind of response is a direct manifestation of the problem
which has been reported in \cite{dir}. Furthermore, \textbf{Figure} 6 also shows that
not only the two steerabilities $\zeta_{m_{1}|m_{2}}$ and $%
\zeta_{m_{2}|m_{1}}$ but also the entanglement $E_{N}^{m-m}$ are strongly
sensitive to the frequencies of the two mechanical modes. Moreover, it can
be seen that when the value of the frequency of mirror-2 is greater than the
mirror-1, the region for the one-way (two-way) steering gets broaden
(shorten). \newline

\textit{One-way (OW) control.} As we discussed, the steerability and
entanglement between the rotational modes are sensitive to their
frequencies. We observe that when the two rotational modes are about to
synchronize, i.e., the ratio of the frequency of the two rotating mirrors
approaches unity, the two-way steering completely vanishes. Furthermore,
\textbf{Figure} 7(a) and 7(d) present a fascinating situation where the states of the
two rotational modes are entangled; though they show steerability in one
direction, which indicates the perfect asymmetry of quantum correlations
(one-way steering). Such an asymmetry of quantum correlations describes that
both Alice and Bob can execute the same Gaussian measurements on their side
of the entangled system, however, they find completely different results.
One-way steering can be explained by the steering asymmetry introduced in
Eq. (\ref{SAS}). \textbf{Figure} 7(a) and 7(d) clearly show that one-way steering
can be observe until $\omega_{\phi_{2}}/\omega_{\phi_{1}}=0.9$ or $%
\omega_{\phi_{2}}/\omega_{\phi_{1}}=1.1$. The one-way steering has been
experimentally realized in \cite{dir,exp1} and imparts one-side device
independent quantum key distribution (QKD). However, we observe no-way
steering when $\omega_{\phi_{2}}/\omega_{\phi_{1}}\simeq0.95$ or $%
\omega_{\phi_{2}}/\omega_{\phi_{1}}\simeq1.05$. Furthermore, when the two
rotating mirrors rotate with exactly the same frequency (complectly
synchronize), even the entanglement between rotating mirrors completely
cease to exist (not shown).

\textit{Effect of Parametric Gain and Phase on quantum steering.} Finally,
the degree of entanglemnt entanglement and dynamics quantum steering can be
studied under influence of the parametric gain for different value of phase
as shown in \textbf{Figure} 8 around the optimum value of the cavity detuning. Firstly,
it can be seen in \textbf{Figure} 8. that quantum steering remains upper bounded by the
entanglement irrespective of which rotational mirror has higher frequency.
\textbf{Figure} 8 (a-d) shows that with gradual increase of the parametric gain, the
two rotational modes exhibit two-way steering except the case when $%
\theta=\pi/2$, where two-way steering converts to one-way steering around $%
\chi/\omega_{\phi_{1}}=0.13$. This is because we obtained reduced
entanglement at this parametric phase (can be seen in \textbf{Figure} 2(a)). The same
situation can also be observe in lower panel of \textbf{Figure} 8, where the one-way
steering vanishes around $\chi/\omega_{\phi_{1}}=0.16$. Owing to the gradual
enhancement of entanglement for $\theta=3\pi/2$, one-way steering converts
to two-way steering for large value of parametric gain. Hence, the current
scheme may provide motivation for handling quantum steering in two
rotational cavity optomechanical system.
%We set $\omega_{\phi_{2}}/\omega_{\phi_{1}}=0.5$ for upper pannel (Fig. 8(a-d)) and $\omega_{\phi_{2}}/\omega_{\phi_{1}}=1.5$ for lowor pannel (Fig. 8(e-h)) and the optimum cavity detuning $\Delta=-\omega_{\phi_{1}}$.
%0.042-0.03

\section{Experimental implementations}

\label{EI} The parameters for the current scheme is given in Table. 1.
Keeping these parameters in mind, we now talk about the feasibility of the
current two rotating system based on the recent experiments. In recent
experiment, the value of the angular momentum is taken very high because it
can easily be realizable via spiral phase elements. In addition, the
azimuthal structure of beam of light can be altered via transmission or
reflection from the spiral phase elements \cite{SYC}. It is shown that low
mass of the rotational mirrors with high precision can be well fabricated.
Futhermore, topological charge of the LG laser beams can be taken as high as
$1000$ \cite{SYC}. With the fast growing nanotechnology, the mechanical
oscillators, with low effective mass ($m=37pg$), high-quality factor ($%
Q_{m}=10^{8}$) and and high frequency (a few MHz), has been experimentally
reported \cite{KHG}. This implies that the current system is an
experimentally feasible system under the present-day technology.

\section{Conclusions}

\label{CON} We have presented an efficient and effective method for
improving the entanglement and one-way (two-way) steering characteristics of
two rotating mirrors in a Laguerre-Gaussian (LG) rotating cavity, which
composed of two rotational mirrors coupled to a cavity via orbital angular
momentum exchange and an optical parametric amplifier (OPA) driven by
coherent light. We have shown that the frequencies of two rotational mirrors
play a vital role in modulating the entanglement and controlling the quantum
steering. For example, when $\omega_{\phi2}=0.5\omega_{\phi1}$ ($%
\omega_{\phi2}=1.5\omega_{\phi1}$), the steerability of two mirrors is
enhanced as compared to the case when the two mirrors are near to
synchronized, that is, near the resonance frequency, two-way steering
converts to one-way steering (or no-way steering) with a reduced degree of
entanglement. It is interesting to mention that the switch from two-way
steering to one-way (or no-way) steering is practicable by changing the frequencies of two rotational mirrors,
parametric gain and phase. Hence, we argued that the steering directivity in the current scheme would
support understanding the quantum correlation and may have major applications in one-way (two-way) quantum computing, quantum secret sharing, and quantum key distribution.
\section*{Data availability}
All numerical data that support the findings in this study is available within the article
%we have shown that the one-way/two-way gaussian
%steering may be achieved in a Laguerre-Gaussian (LG) rotating cavity optomechanical system, which
%Finally, one-way/two-way gaussian steering may be achieved in a Laguerre-Gaussian (LG) rotating cavity optomechanical system and may find major applications in one-way (two-way) quantum computing, quantum secret sharing, quantum key distribution, etc.
%Therefore, our scheme provides inspiration for manipulating the asymmetric quantum steering with enhanced steerability via a
%Our present scheme is realized with two YIG spheres coupled to a microwave cavity
%Our scheme shows one advantage: the system creates one-way steering without imposing additional conditions of asymmetric
%losses or noises in the subsystems, but changes the population of the two magnon modes through the ratio of coherent
%information exchange frequencies.
%We have demonstrated an effective and a straightforward approach to $\varphi$

\end{document}